# Autonomous robotic mechanical exfoliation of two-dimensional semiconductors combined with Bayesian optimization

*Fan Yang[1, 2, 3, 4†], Wataru Idehara[4†], Kenya Tanaka[4], Keisuke Shinokita[4],*

*Haiyan Zhao[1, 2, 3]*, and Kazunari Matsuda[4]**

[1]Department of Mechanical Engineering, Tsinghua University, Beijing 100084, China

[2]State Key Laboratory of Tribology in Advanced Equipment, Tsinghua University, Beijing 100084, China

[3]State Key Laboratory of Clean and Efficient Turbomachinery Power Equipment, Department of Mechanical Engineering, Tsinghua University, Beijing 100084, China

[4]Institute of Advanced Energy, Kyoto University, Uji, Kyoto 611-0011, Japan

[†]equal contributin for this study
*corresponding author: Haiyan Zhao, Kazunari Matsuda
E-mail: hyzhao@tsinghua.edu.cn, matsuda@iae.kyoto-u.ac.jp

## Abstract

Simple mechanical exfoliation of layered materials is the most frequently employed method for producing high-quality monolayers of two-dimensional semiconducting materials. However, the mechanical exfoliation by human hands is a microscopically sophisticated process with a large number of microscopic parameters, which requires significant operator efforts and limits the reproducibility in achieving high-quality and large-area monolayer semiconducting materials. Herein, we have proposed a new strategy for the mechanical exfoliation by combining a developed robotic system and Bayesian optimization. We have demonstrated that it is possible to explore the optimized experimental conditions among a large number of parameter combinations for mechanical exfoliation in a relatively small number of experimental trials. Moreover, the entire mechanical exfoliation process from preparation to detection of monolayer semiconductors was performed by the developed autonomous robotic system. The optimized experimental condition was determined through only 30 trials of mechanical exfoliation experiments, representing 2.5% of all experimental parameter conditions. As a result, the critical parameters for the efficient fabrication of large-area monolayer $WSe_2$ were elucidated.

## Introduction

Recently, atomically thin two-dimensional (2D) materials[1] including graphene, and monolayer transition metal dichalcogenides have attracted much attention in a variety of disciplines due to their electronic and optical properties that do not appear in their bulk crystals. The monolayer transition metal dichalcogenides ($MX_2$:M=Mo, W, X=S, Se) are therefore highly anticipated as the next-generation semiconductors for potential applications of electrical logic circuits, light emission/detection, quantum light source[3], solar cells[4] and so on[5,6]. Several methods for fabricating monolayer 2D semiconductors with thicknesses of only a few nanometers are mechanical exfoliation, chemical exfoliation[7], liquid phase exfoliation[8], chemical vapor deposition[9], and pulsed laser deposition[10]. Among these methods, the mechanical exfoliation from bulk single crystals has been widely and frequently employed to fabricate the high quality graphene and monolayer 2D semiconductors for the state-of-the-art research[11, 12]. The high-quality graphene and monolayer 2D semiconductors fabricated by the mechanical exfoliation support the emerging novel phenomena such as superconductivity of bilayer graphene, optical phenomena arising from moiré excitons and high-performance field-effect transistors[13-17].

Nevertheless, the mechanical exfoliation method itself is recognized as a simple process; however, it encounters significant bottlenecks due to the substantial manpower requirements. The process is composed of numerous steps, such as substrate cleaning, exfoliation of bulk single crystals, transfer of small flakes to the substrate, and monolayer detection, which make it challenging to efficiently prepare the large-area monolayer 2D materials over 100 μm². Moreover, the experimental conditions must be carefully selected from a vast number of potential parameters such as types of tape, folding time, peeling velocity, and so on, and the detailed microscopic mechanism of mechanical exfoliation itself has yet to be elucidated. At the present stage, the only ways are to wait for serendipity by repeating the many experimental trials, or to rely on experienced and skilled resaerchers to find large graphene and monolayer 2D materials. Recently, several advanced methods have been propsed to support the fabrication process by using machine learning to identify monolayer 2D materials[18,19] and robots to automatically stack exfoliated 2D materials[20]. These efforts represent the next generation of research aimed at combining robotic systems and machine learning to efficiently advance all research in the field of 2D materials on a large scale. In this context, it is strongly required to develop the efficient and highly reproducible strategy to fabricate the large-area and high-quality monolayer 2D materials by mechanical exfoliation[21, 22], which would provide a significant impact on the wide range of research area in 2D materials fundamental research and development.

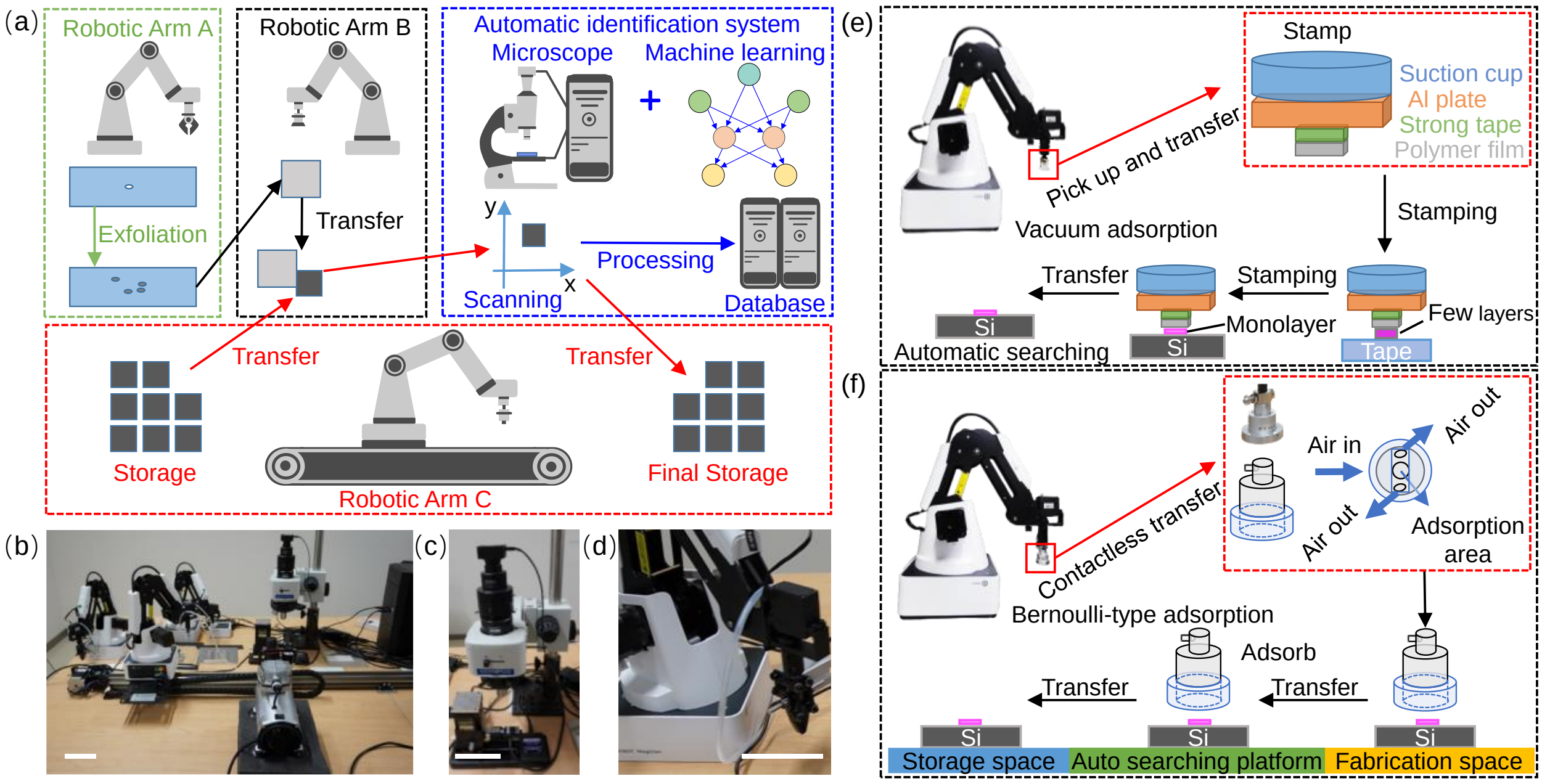

**Figure 1 Schematics, functionalities, and photographs of the developed robotic system.** (a) Schematic representation of the 2D monolayer exfoliation and searching process. Initially, the single crystals of 2D materials on the blue-tape are repetitively folded by the robotic arm-A. Subsequently, the small flakes of 2D materials are transferred to the Si substrate by the designed stamp of the robotic arm-B. An automatic detection system based on machine learning is employed to identify the 2D monolayers and catalogue their position in a database. Subsequently, the Si substrate with monolayer 2D materials is storaged. (b-d) Photographs of (b) the entire system, (c) the optical automatic detection and identification platform, and (d) the blue-tape folding apparatus. The scale bars in the photographs correspond to 10 cm. (e) Schematics of the automatic stamping system with the suction cup attachment of robotic arm-B. (f) Schematics of automatic transfer system with the customized Bernoulli-type adsorption component of robotic arm-C.

Herein, we have demonstrated the autonomous robotic mechanical exfoliation of 2D semiconductors supported by artificial intelligence (AI) algorithms including machine learning (ML)[23-27] and Bayesian optimization (BO)[28-32]. The developed robotic mechanical exfoliation system eliminates the need for human labor and reduces the uncertainty of results, which enables the stable fabrication of high-quality, and large-area monolayer semiconductors. The Bayesian algorithm accelerates the optimization of mechanical exfoliation among the huge number of experimental conditions. The integration of the robotic system and Bayesian optimization will facilitate the development of strategies for maximizing the 2D materials research.

## Results and Discussion

### System architectures and functionalities

**Figure 1a** illustrates the schematics of the developed robotic fabrication and searching system for high-speed and reproducible mechanical exfoliation and detection of monolayer 2D semiconductors. The system is composed of robotic mechanical exfoliation, transfer from the mechanical exfoliation to detection, and automatic searching and identification of monolayer 2D semiconductors equipped with a microscope. **Figures 1b-d** show the optical images of the whole robotic system we have developed, the optical automatic identification platform, and the folding device, respectively.

In the robotic system, the high-precision robotic arm-A first repeatedly folds the blue-tape for the mechanical exfoliation on which the single crystal flakes have been pre-positioned. During the exfoliation process, the size and thickness of crystal flakes gradually decrease, similar to that by manual mechanical exfoliation by human hands as reported previously. Once the blue-tape is sufficiently folded, the robotic arm-C transfers a cleaned silicon (Si) substrate to the exfoliation platform. The robotic arm-B then pushes and transfers the small 2D flakes from the tape surface to the Si substrate using a specially designed stamp. The robotic arm-C, equipped with a customized Bernoulli-type adsorption component, then transfers the Si substrate containing many small flakes to the optical auto-identification platform. In the identification system composed of the optical microscope with a CMOS camera and a high-speed motorized XYZ scanning stage, the system can automatically focus, scan the surface of the Si substrate, and capture optical images of the Si substrate surface with many small flakes including monolayers. The captured optical microscopy images are analyzed using computer vision and machine learning algorithms to determine the presence of monolayer semiconductors on the Si substrate in each image. When the flakes of monolayer are detected, their location and morphology are cataloged in a database. Finally, the robotic arm-C transfers the inspected Si substrate to a designated storage location.

**Figure 1e** shows the schematics of the automated stamping system, integrated with a suction cup attachment in the robotic arm-B, which is used to transfer the small flakes including monolayers from the exfoliated blue-tape to the Si substrate. The transfer of monolayer flakes from the tape to the substrates is a major challenge in the fabrication of monolayer 2D materials[33]. This process relies on manual intervention, which makes it difficult to control the pressure and contact area between the tape and the substrates during the transfer process, resulting in significant uncertainty and reprodicibility in the process. To solve this problem, we developed a customized stamp consisting of a suction cup, iron plate, strong adhesive tape, and polymer film, which effectively facilitates the transfer of exfoliated flakes from the blue-tape to the Si substrate. The stamp peeling method is applicable to transition metal dichalcogenides, graphene, and hexagonal boron nitride (*h*-BN)[34,35]. The robotic arm-B slowly presses the stamp onto the folded blue-tape with small flakes of 2D materials, ensuring contact between the stamp and the blue-tape to transfer the flakes onto the stamp. The contact speed between the stamp and the blue-tape is controlled by the operating program of the robotic arm-B at typically 1 mm/s to avoid damage to the flakes. After sufficient pressing cycles, the stamp is slowly lifted, consequently transferring the many small flakes from the blue-tape to the polymer film.

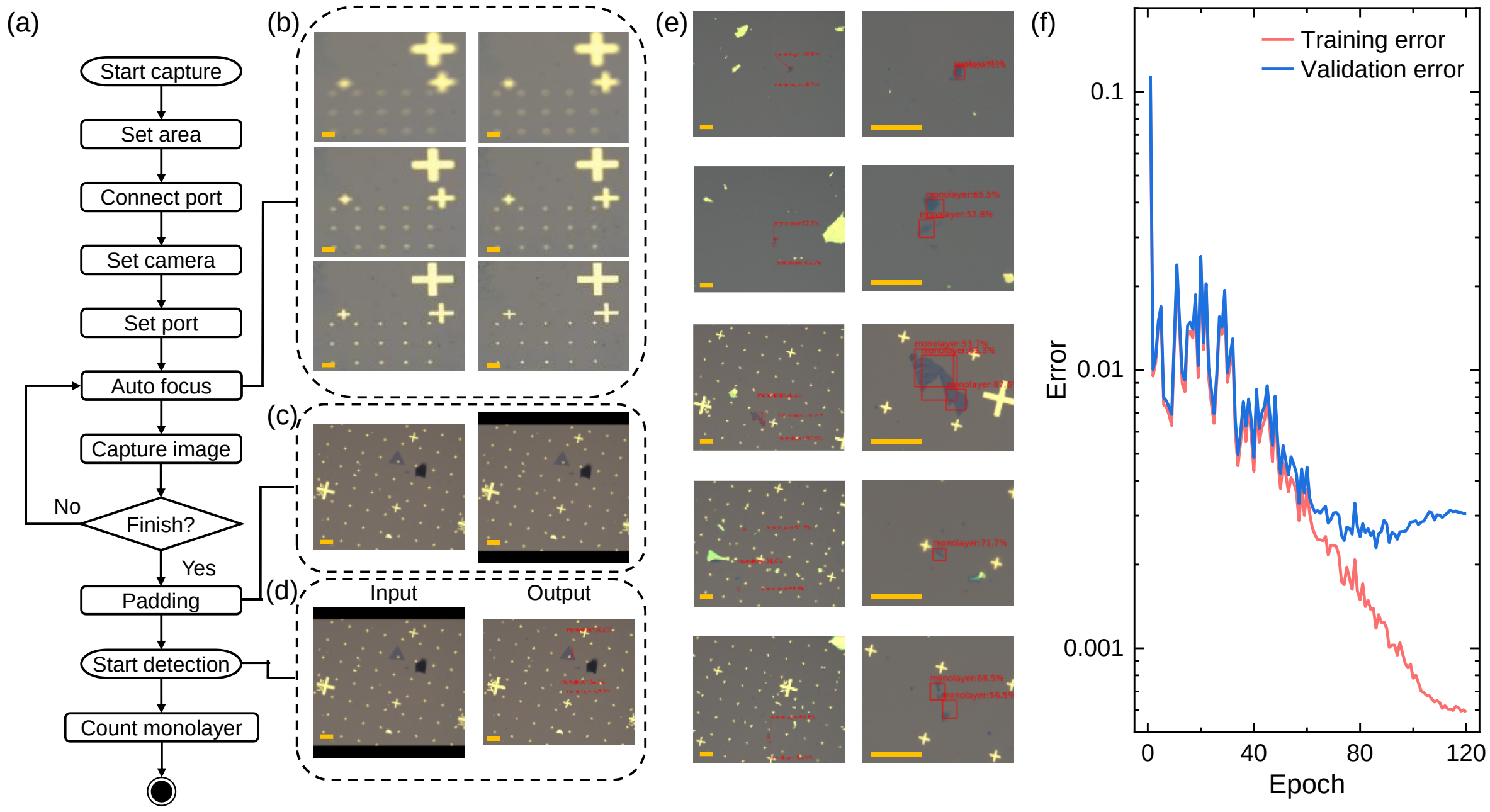

**Figure 2 Flowchart of automatic monolayer identification algorithms.** (a) Schematic of identification algorithms for automatic capture of optical microscope images and the recognition of monolayer 2D materials. (b) Schematic of the autofocusing procedure for the image capture system. (c) Padding for machine learning datasets. (d) Application of machine learning algorithms to the delineation of bounding boxes. (e) Left side section is as follows: The optical images of $WSe_2$ were padded by the automatic identification system equipped with a microscope. The right section displays: The results of machine learning-based monolayer detection process are presented. The scale bars in each image corresponds to a length of 40 μm. (f) Learning curves for monolayer identification. The red and blue line represents the training error, and the validation error, respectively. A total of 120 learning epochs were conducted. Both training and validation errors systematically decrease with the increase of learning epoch.

The stamp is brought into contact with the cleaned Si substrate by controlling the robotic arm-A, repeating a similar contact process between the film and small flakes to the Si substrate, as described above. Finally, the flakes with various thick materials such monolayer, bilayer, and thick bulk are transferred onto the surface of the Si substrate by van der Waals (vdW) force between the flakes and the Si substrate. The details of procedures and conditions of the stamp and 2D flake exfoliation are described in the Methods section. **Figure 1f** shows schematics of automatic transfer system with a customized Bernoulli-type adsorption component on robotic arm-C. The adsorption component in the robotic arm-C is used to achieve contactless pickup and transfer of the Si substrate by Bernoulli-type air flow, which is realized by the pressure difference between the top and bottom surfaces of the substrate, to avoid the contact and resultant contamination of the surface of 2D material flakes. Moreover, a specially designed attachment is employed to the Bernoulli-type adsorption component to avoid the rotation of Si substrate and keep the position of Si substrate within the accuracy of 0.1 mm during the process.

## Identification algorithms to capture images and recognize 2D monolayers

**Figure 2a** illustrates the workflow of the automatic monolayer detection system, which is based on machine learning[36] by utilizing the Neural Network Console (NNC) provided by Sony[37]. An optical image is captured using a 20×objective lens with a simultaneous autofocusing process (also see in **Figure 3a**). It takes only typically 20 min to acquire a large number of autofocused optical images of the surface in each Si substrate with a 9×10 mm. The original images for the dataset were expanded to 1024 pixels in both width and height to improve the accuracy of machine learning algorithms (also see in **Figure 3b**). To prepare an image dataset for machine learning, the labeling annotation tool was first used to delineate the boundaries of monolayer flakes in the optical images. In this case, three labels are categorized : "monolayer", "thin", and "bulk" are assigned value of 0, 1, and 2, respectively. A text file in YOLO[38] format is generated containing information on the center position, width and height of each object. The large number of acquired optical images and the corresponding text files were imported into NNC, and the machine learning algorithm was configured.

**Figure 2f** shows the learning curve of a machine learning algorithm[38] for automated monolayer flake detection. We employed the dataset consisted of 960 optical images of $WSe_2$ flakes on the Si substrate; 70% of the dataset were used for training and the remaining 30% for validation. Both the training error (red line) and validation error (blue line) in **Figure 2f** clearly show a systematic decrease with each epoch, which indicates that the learning process is successful. The detected bounding box almost completely encloses the monolayer flakes with a successful detection rate exceeding 95%, which makes it an excellent algorithm for the practical use of monolayer detection. The previous automatic monolayer detection system using only RGB information (see in **Supplementary Fig. S1, S2**), was unable to effectively distinguish monolayer flakes from the contamination of blue-tape, and so on. Moreover, the machine learning detection system eliminated the need to manually adjust the threshold each time based on the ambient brightness, resulting in significant labor savings. The optical images were then stored in a separate folder and the number of monolayer flakes and their areas were automatically counted for further analysis described below. We confirmed that the detected $WSe_2$ and $MoSe_2$ flakes fabricated by the robotic fabrication and automatic detection system have monolayer thickness by photoluminescence (PL) and Raman scattering spectroscopy (see in **Supplementary Fig. S3**), which are consistent with those of the manually prepared and detected monolayer flakes by human hands, and eyes under the microscope.

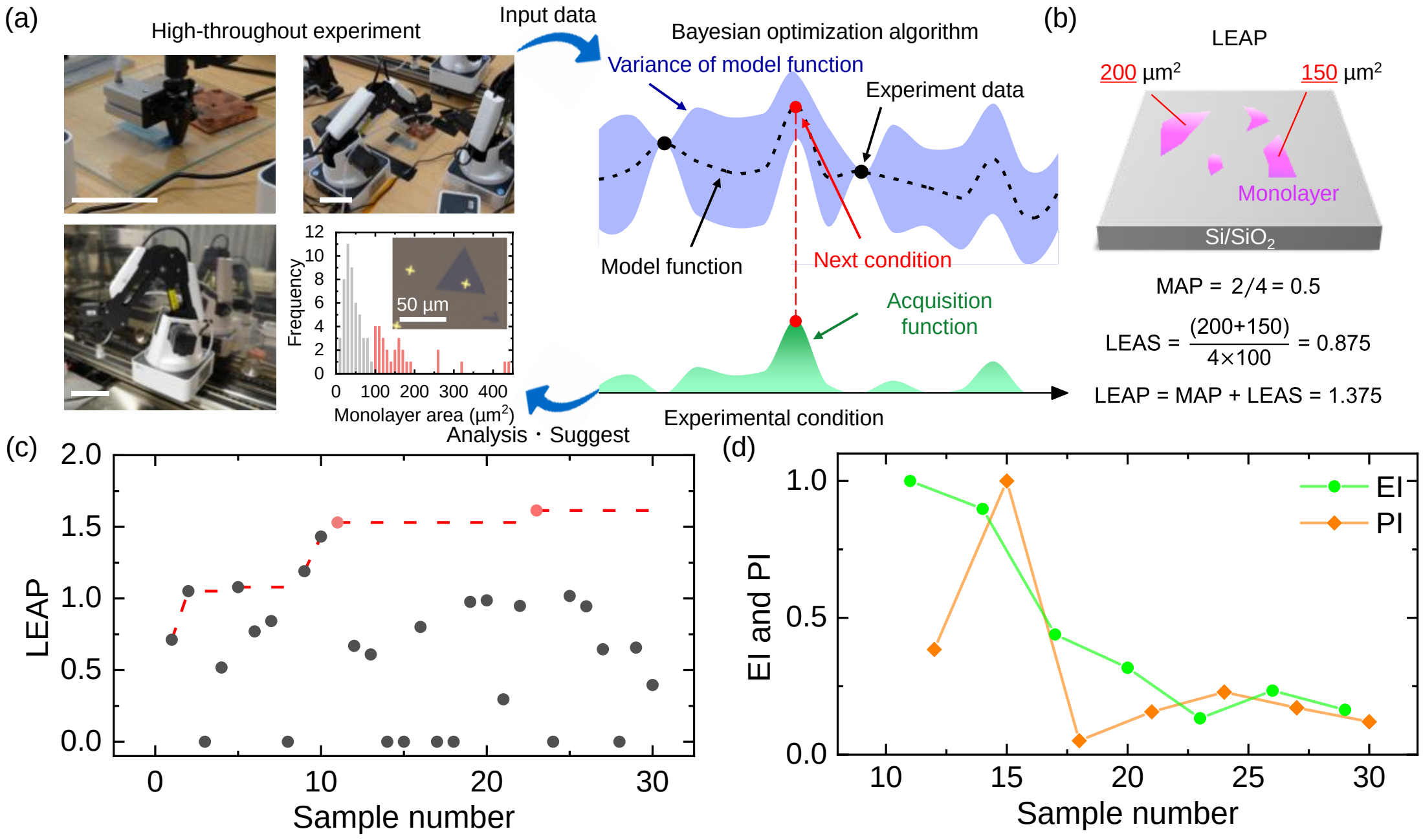

**Figure 3 Index of mechanical exfoliation and Bayesian optimization.** (a) Overview of Bayesian optimization combined with the developed robotic system. The scales in each image show a length of 10 cm. (b) Definition of index for mechanical exfoliation as large exfoliated area performance (LEAP) to maximize the Bayesian optimization. An example of LEAP formula is shown. (c) Evolution of updating the maximum score of LEAP in all 30 Bayesian optimization experiments. (d) Evolution of acquisition functions of expected improvement score (EI) and predicted improvement score (PI) over all 30 experiments, where the EI and PI are normalized.

## Bayesian optimization

We attempted to efficiently explore the optimized experimental conditions for mechanical exfoliation by combining the developed robotic system with Bayesian optimization (BO), because the mechanical exfoliation is a simple process; however, it involves a huge number of experimental conditions. In BO, it is necessary to set parameters that have a significant impact on the results in order to efficiently explore a large parameter space, and the parameter space was set based on experiments conducted with the developed robotic system. Moreover, we systematically varied and selected the parameters that had a greater effect on the results of mechanical exfoliation. The type of blue-tape, number of blue-tape folding, peeling velocity, and number of transfer onto the blue-tape were selected, resulting in a total of 12,000 experimental conditions. Such a huge number of experimental conditions makes the tasks difficult for optimization without the support of data science approaches such as BO.

Several initial conditions should be provided to construct an appropriate model function for the BO algorithm [39,40]. The initial ten experimental conditions are selected from all the experimental conditions (12,000 experimental conditions) using the D-optimization criterion[41] to ensure that the characteristics among the parameters are not constant. Considering the results of these initial conditions, the appropriate kernel function[42] is used to construct the model function. Moreover, the acquired number and integrated total area of monolayer flakes were treated as indicators. However, these are not suitable for the BO algorithm due to their high variance even under the same experimental conditions (see in **Supplementary Fig. S4**). A new index that indicates the quality of mechanical exfoliation results needs to be designed according to the target, i.e., efficiently acquiring many monolayer flakes larger than 100 $\mu m^2$ suitable for the fabrication of devices and vdW heterostructures.

**Figure 3a** shows the experimental flow of BO combined with the robotic system. The experiments of mechanical exfoliation are mainly performed by the developed robotic system without human intervention. The obtained results are input to the BO algorithm, which constructs the model function, predicts the variance range, and calculates the acquisition function. The expected score that updates the maximum in the probable experimental condition is suggested by the type of acquisition function. Then, the next experimental condition according to the predicted parameters by BO is performed. These series of epochs are repeated until a sufficiently high score is obtained or the expected score and the maximum update probability reach physical limits. **Figure 3a** also shows the histogram of monolayer $WSe_2$ as a function of area in the optimized experimental conditions obtained in the experiments after 8 iterations. The results indicate that the number of acquired monolayer $WSe_2$ flakes by mechanical exfoliation follows a normal distribution. An index that maximizes the red parts with a large area in the histogram of **Figure 3a** is desirable. Then, we introduced the new evaluation index as a large exfoliated area performance (LEAP).

**Figure 3b** shows a schematic of LEAP calculation as an index for BO. The LEAP is an index that indicates how large and many monolayer $WSe_2$ can be efficiently obtained with a threshold areal size of 100 $\mu m^2$. The LEAP is composed of the value of monolayer appearance probability (MAP: the percentage of monolayers produced that exceed 100 $\mu m^2$) and large exfoliated area score (LEAS: the average size of monolayers that can be fabricated), corresponding to indices for the number and area of monolayer flakes, as described beow,

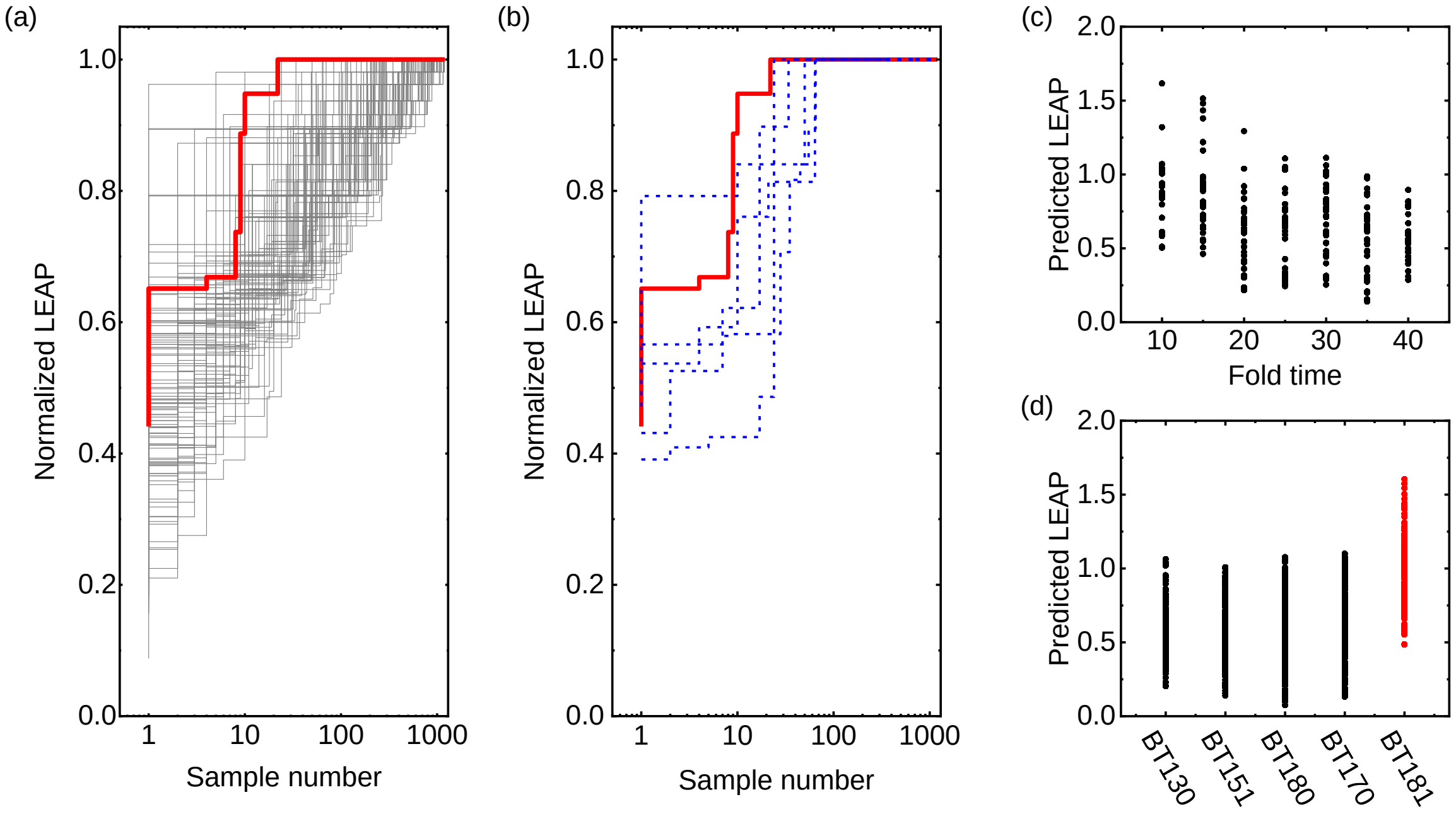

**Figure 4 Convergence performances in Baysian optimization and randam simulation.** (a) Comparison of convergence performance between Bayesian optimization and random simulation. Grey lines and red line show the simulated traces by the random simulation, and the Baysian-optimization, respectively. (b) Blue lines show the fastest five traces that reached to convergence in the random simulations. The score of LEAP are normalized to compare. (c) and (d) Predicted LEAP as function of the folding time and type of blue-tapes. The plot also shows the distribution of LEAP when the parameters are fixed, based on the results of random simulation.

$$\mathrm{LEAP} = a_1 \frac{n}{N} + a_2 \frac{S}{N \times 100} \qquad (1)$$

where the first term and second term correspond to MAP and LEAS, and $a_{1(2)}$ are coefficient of MAP and LEAS, respectively, and $N$ and $n$ indicate the total number of $WSe_2$ monolayers larger than 10 μm² and 100 μm² , respectively. Also, $S$ indicates the total area of $WSe_2$ monolayers larger than 10 μm². We can optimize the mechanical exfoliation process from the optimal balance between number and area of monolayer flakes using the index of LEAP in BO procedures. These MAP and LEAS coefficients need to be properly set, so that the variance of LEAP is reduced. The coefficients $a_{1(2)}$ and monolayer area thresholds for each of the MAP and LEAS parameters are examined in **Supplementary Fig. S5 and S6**. As a result, the introduction of LEAP index largely contributes to the reduction of variance in the histogram (**Supplementary Fig. S5**) in each dataset, which enables the BO to proceed in the mechanical exfoliation.

**Figure 3c** shows the evolution of the evaluated LEAP in each experimental trial, corresponding to how the LEAP updates its maximum score in 30 BO trials as the number of trials increases. In the first 10 experimental trials, the BO algorithm was able to update the maximum score, indicated by red circles in **Figure 3c**, over the next 11 experimental conditions, and the maximum score is not updated after the 23rd experimental trial. In other words, within 30 trials, BO is able to find the optimized experimental conditions among the first 22 trials. Note that the BO algorithm depends on the acquisition function used and does not explore the entire parameter space. The three acquisition functions used in this experiment were EI (expected improvement score), PI (probability of improvement score), and MI (mutual information score), each applied in turn. MI is the acquisition function that primarily proposes unexplored regions and suggests experimental conditions in which the predictions plus variation from the model function are maximized. In **Figure 3d**, we show the trends for EI and PI in order to understand whether we are searching more efficiently. We can see that the acquisition functions[43] for EI and PI clearly decrease as the number of experimental trials increases. Moreover, both EI and PI decrease sharply around 15 experimental trials, which is well consistent with the saturation of updated behaviors after 15 experimental trials in the LEAP score, as shown in **Figure 3c**. These results indicate that the optimized conditions are efficiently explored as the number of experimental trials increases. Moreover, the value of EI is significantly lower than 1.15, corresponding to hypothetically adding a monolayer flake of 100 μm$^2$, at the time of the 23rd experimental trial. In other words, it is unlikely that the maximum LEAP score will be significantly updated beyond the 23rd experiment trial. Maximizing the LEAP index leads to more efficient fabrication of many large monolayer flakes over 100 μm$^2$, corresponding to the red parts of the histogram in **Figure 3a**.

## Comparison between BO and random simulation

We simultaneously compared the results of BO with predicted random simulation in order to show the effectiveness of BO in the mechanical exfoliation process. **Figure 4a** shows a simulated benchmark of calculated LEAP by selecting the parameters in the framework of random simulation and BO. In the random simulation, 1,200 randomly selected conditions are chosen from all experimental conditions of 12,000, and 100 patterns are prepared to compare how quickly the maximum score of LEAP is reached to the score derived in BO. The predicted score in the random simulations is calculated based on a model function constructed in BO. The results of cross-validation used to construct the model function are shown in **Supplementary Fig. S7**. Moreover, the maximum LEAP score in each simulation pattern is normalized to compare the number of trials required to reach the maximum score. The LEAP in BO reaches the maximum score after only 23rd trials, as shown in **Figure 4a**, while the LEAP in random simulation reaches the maximum score after an average of 485 trials. **Figure 4b** shows the evoluation of predicted LEAP in BO and random simulation for the selected fastest five patterns, as a function of trials. These results clearly show that the Bayesian algorithm is much more effective and a superior optimization than the random algorithm for the optimization of mechanical exfoliation process.

**Figures 4c and 4d** show the predicted LEAPs as a function of folding times of blue-tape and types of blue-tape, respectively, where the folding time and types of blue-tape are found to be important parameters in the mechanical exfoliation process. Other experimental parameters are shown in **Supplementary Fig. S8**. And the detailed properties of blue-tapes are shown in **Supplementary Table S1**. Accordingly, it was found that the highest predicted LEAP score depends on the thickness of the adhesive layer of the blue-tape[44], as shown in **Figure 4c**, rather than the adhesive strength of the blue-tape surface, as shown in **Supplementary Table S1**. We also found that the predicted LEAP score decreases as the number of the folding times increases in **Figure 4d**. This result indicates that the smaller number of the folding times of blue-tape is better regardless of other parameters, and that the fewer folding time results in a larger area of thin 2D flakes remaining unbroken on the tape surface. Finally, a large number of monolayer flakes with a large area over 100 μm$^2$ are obtained on each Si substrate in the BO process. These results suggest that a robotic system using the BO algorithm for mechanical exfoliation would be useful for efficient fabrication of many other monolayer 2D materials, without human experiences and large number of experimental trials.

## Conclusion

In summary, we have demonstrated a new strategy for mechanical exfoliation of monolayer semiconductors by combining Beyesian optimization and an autonomous robotic system. The developed robotic system is capable of consistently executing the whole process from fabrication to detection of monolayer flakes, thereby eliminating the need for human labor and reducing the uncertainty of results, which enables the stable fabrication of high-quality and large-area monolayer semiconductors. Moreover, developed robotic mechanical exfoliation and autonomous detection with high reproducibility also enable us to employ the Bayesian algorithm for the optimization of mechanical exfoliation among the huge number of experimental conditions. We have demonstrated the effectiveness of BO for the mechanical exfoliation of monolayer semiconductors, in which the mechanical exfoliation to fabricate a large number and large area of monolayer flakes can be realized in only 30 experimental trials among 120,000 experimental trials. These results show a much faster convergence than that in the case of random simulation. These results of this study would provide a new strategy for a mass-production platform for the fabrication of monolayer 2D semiconductors and potential applications in the 2D materials research and development.

## Methods

## Hardware and software design

We have integrated a series of hardware components and developed the software system to meet the requirements of automated exfoliation and identification of monolayer 2D materials. The three desktop collaborative robots with the four-axis robotic arms (Dobot Magician, Dobot Robotics) are used as robotic arm-A, -B, and -C. The gripper attachment on the robotic arm-A is used to perform the folding process of the blue-tape, and the suction cup attachment on the robotic arm-B is used to transfer the 2D flakes from the blue-tape to the Si substrate. The robotic arm-C on the conveyor facilitates the transfer of the Si substrate to the automatic detection system. The operating program is written in both Python and Matlab, where the Python program controls the robotic system and the Matlab program serves as a control terminal and executes image identification algorithms. The entire system is elaborately designed to automate the mechanical exfoliation, search, detection, and storage of monolayer 2D materials by simply entering a few instructions.

## Customized stamp construction

The desinged stamp is composed of a suction cup, an aluminum (Al) plate, and a polymer film. The designed stamp for transferring the flakes of 2D materials to the surface of Si substrate is picked up by the suction cup of the robotic arm-B. The Si substrate is fixed by a vacuum pump at the base of the exfoliation platform during the exfoliation process. The vacuum pump is controlled by a remotely programmable smart plug (SwitchBot), which allows precise activation and deactivation, so that the Si substrate remains fixed during the transfer process of flakes from the polymer film to surface of Si substrate. Then, the Si substrate with many flakes is released from the exfoliation platform, picked up by the robotic arm-C, and transferred to the monolayer detection system.

## Machine learning

The machine learning for monolayer detection was performed using the application platform of Neural Network Console provided from Sony[36]. The acquired optical images are formatted in YOLO[38] format. The lower left corner of each optical image is set to the coordinate $(x, y) = (0, 0)$, and the object is output in the format (label of the object, relative width of the object, relative height of the object). After formatting the obtained optical images in the YOLO, we loaded them as a dataset into NNC. The training was performed using a sample program of NNC "Object Detection". The program requires setting the image width, height, and number of anchors, appropriately. To use the GPU for training, it is necessary to set the NVIDIA GPU and corresponding cuda folder path in the NNC engine tab. A code has been written using Python to draw the bounding box in the optical image, based on the sample code[37]. All these process are opened on Github.

## Bayesian optimization

We used a sample program opened on GitHub (https://github.com/hkaneko1985/python_doe_kspub) as a program to perform Bayesian optimization[45], as follows.

1. Preparation (generating random parameter sets)

We set the upper and lower bounds of each parameter and created 12,000 random parameter sets, using "sample_program_05_01_sample_generation.py".

2. Set initial conditions

The 10 parameter sets are randomly selected from the previously created parameter sets and generated using the D-optimal criterion, by "sample_program_05_02_sample_selection.py". The D-optimal criterion is an efficient method that minimizes $\det((X'X)^{-1})$ and has the least confusion of model terms. It is possible to create parameter sets without overlapping information in this scheme.

3. Bayesian optimization

According to the initial conditions, the mechanical exfoliation of bulk $WSe_2$ single crystals was performed for four Si substrates per experimental trial, and the data were averaged to reduce the statistical errors. The score of LEAP was calculated based on the number and area of detected monolayer flakes. The score of LEAP in each experimental condition was saved in the input file, which is then entered into "sample_program_05_04_bayesian_optimization.py". We executed calculations for the acquisition function and the program output the parameters for the next mechanical exfoliation condition. The acquisition functions were EI (expected improvement score), PI (probability of improvement score), and MI (mutual information score), respectively.

## Data availability

Data presented in this paper and the supplementary materials are available from the corresponding author upon request.

# Supplementary Information

## Autonomous robotic mechanical exfoliation of two-dimensional semiconductors combined with Bayesian optimization

*Fan Yang[1, 2, 3, 4†], Wataru Idehara[4†], Kenya Tanaka[4], , Keisuke Shinokita[4],*

*Haiyan Zhao[1, 2, 3]*, and Kazunari Matsuda[4]**

[1]Department of Mechanical Engineering, Tsinghua University, Beijing 100084, China

[2]State Key Laboratory of Tribology in Advanced Equipment, Tsinghua University, Beijing 100084, China

[3]State Key Laboratory of Clean and Efficient Turbomachinery Power Equipment, Department of Mechanical Engineering, Tsinghua University, Beijing 100084, China

[4]Institute of Advanced Energy, Kyoto University, Uji, Kyoto 611-0011, Japan

[†] equal contribution for this study

*corresponding author: Haiyan Zhao, Kazunari Matsuda,

E-mail: hyzhao@tsinghua.edu.cn, matsuda@iae.kyoto-u.ac.jp

**Supplementary Note Fig. S1. Detail analysis of flakes obtained from the automated robotic system**

Fig. S1a presents the size distribution of the exfoliated monolayer $WSe_2$, which shows the normal-distribution. To further obtain more information, we also calculated the correlation between the number of monolayers and the bulk materials on the Si substrate. The parameters related to bulk material were calculated as follows:

$$f_{\mathrm{mask}}(x,y) = \begin{cases} 1, & M^{\mathrm{R,G,B}} \le P^{\mathrm{R,G,B}}(x,y) \le N^{\mathrm{R,G,B}} \\ 0, & \text{otherwise} \end{cases} \tag{S1}$$

where $(x, y)$ is the coordinates of a pixel in the image, and $M^{\mathrm{R,G,B}}$ and $N^{\mathrm{R,G,B}}$ are the maximum and minimum values in the histogram for red, green, and blue color component, gained manually prior to initiating the automated recognition.

$$I^{\mathrm{R,G,B}} = \sum f_{\mathrm{mask}}(x,y) \tag{S2}$$

$$I_{\mathrm{R}}^{\mathrm{R,G,B}} = I_{\mathrm{D}}^{\mathrm{R,G,B}} - I_{\mathrm{BG}}^{\mathrm{R,G,B}} \tag{S3}$$

where $I_{\mathrm{D}}^{\mathrm{R,G,B}}$, $I_{\mathrm{BG}}^{\mathrm{R,G,B}}$ and $I_{\mathrm{R}}^{\mathrm{R,G,B}}$ are the bulk threshold color parameter of whole auto-searching image, background, and bulks. Each image is examined one by one via the following equation:

$$T_{\mathrm{B}} = \begin{cases} n+1, & I_{\mathrm{R}}^{\mathrm{R,G,B}} > 0 \\ n, & \text{otherwise} \end{cases} \tag{S4}$$

where $T_{\mathrm{B}}$ is the bulk amount for pictures captured by the automatic identification system.

In a Si substrate, the bulk density is calculated by:

$$\rho_{\mathrm{B}} = \frac{\sum I^{\mathrm{R,G,B}}}{T} \tag{S5}$$

$$\rho_{\mathrm{BT10}} = \frac{\sum_{i=1}^{10} I^{\mathrm{R,G,B}}}{10} \tag{S6}$$

$$\rho_{\mathrm{BT20}} = \frac{\sum_{i=1}^{20} I^{\mathrm{R,G,B}}}{20} \tag{S7}$$

$$\rho_{\mathrm{BT50}} = \frac{\sum_{i=1}^{10} I^{\mathrm{R,G,B}}}{50} \tag{S8}$$

where $T$ is the total number of pictures captured by the automatic identification system. Then, $\rho_{\mathrm{B}}$ is the bulk density of whole Si substrates; $\rho_{\mathrm{BT10}}$, $\rho_{\mathrm{BT20}}$ and $\rho_{\mathrm{BT50}}$ are bulk density in the top 10, 20, and 50 images with the highest density, respectively.

Fig. S1b-f illustrate the relationship between monolayer yield and various bulk parameters. These results indicate that the number of monolayers is roughly positively correlated with these parameters. In other words, we can enhance the monolayer yield by increasing the bulk density on the Si substrate, which provides a direction for the future research on autonomous robotic mechanical exfoliations of 2D materials.

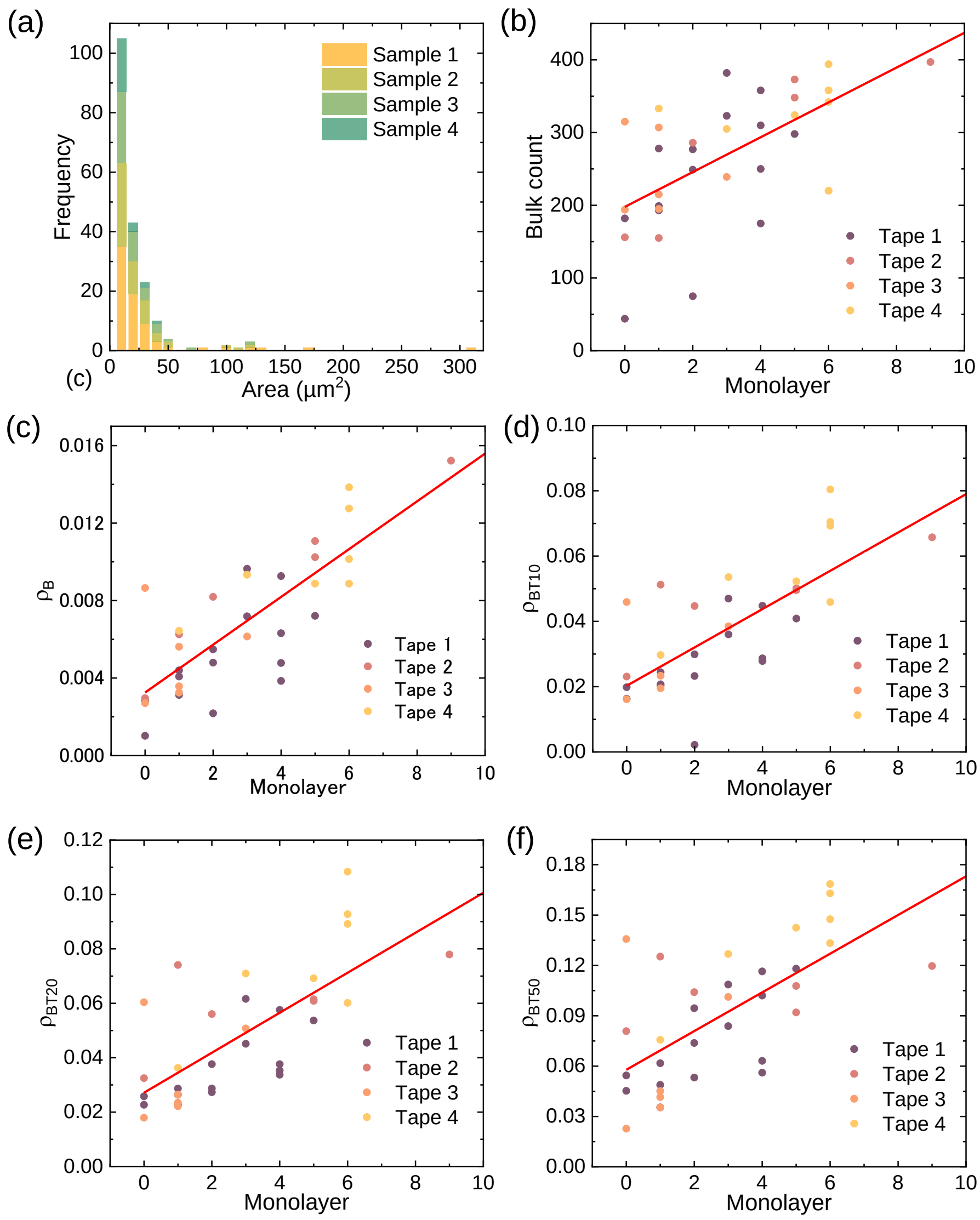

**Figure S1 Analysis of robotic mechanical exfoliation.** (a) Histogram of mechanically exfoliated monolayer $WSe_2$ as a function of area. (b-f) The relationship between monolayer yield and (b) bulk amount, (c) bulk density, (d)-(f) bulk density in top 10, 20, and 50 images with the largest density. These demonstrates that the quantity and dimensions of the bulk crystals, in addition to the number of monolayers, are correlated.

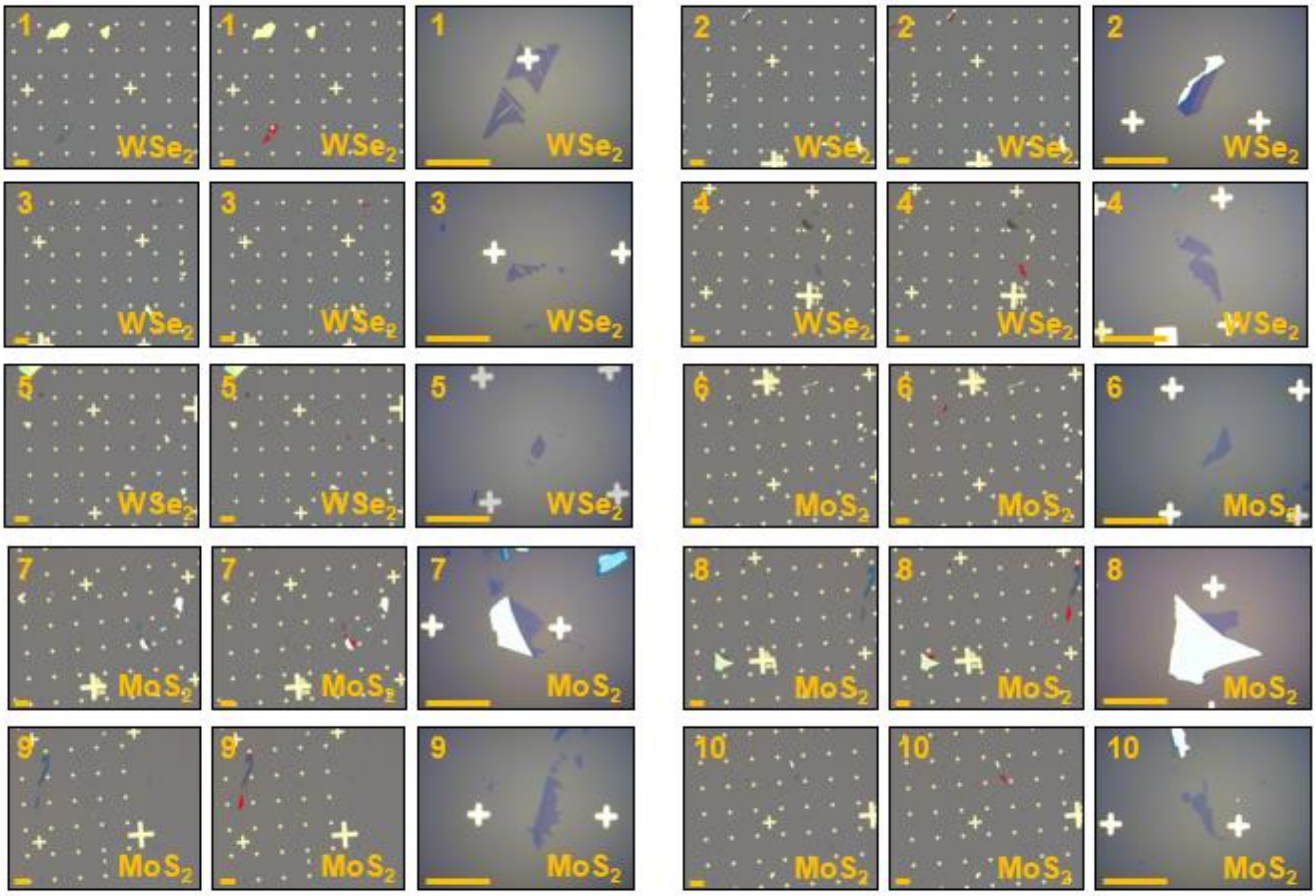

**Figure S2 Results of automatic monolayer identification algorithms.** Left section: Optical images of $WSe_2$ and $MoS_2$ on the Si substrate taken by the microscope of automatic identification system. Middle section: Red color-filled areas correspond to monolayer $WSe_2$ and $MoS_2$ extracted from the monolayer identification algorithms of RGB information. Right section: Optical images of monolayer $WSe_2$ and $MoS_2$ magnified by a 100x objective lens. The scale bar of 30 μm is shown in each image.

**Supplementary Note Fig. S3. Optical measurements of monolayer $WSe_2$ and $MoS_2$**

Monolayer (1L) $WSe_2$, and $MoS_2$ were exfoliated from bulk single crystal and transferred on 270 nm $SiO_2$/Si substrates by the robotic mechanical exfoliation system. A semiconductor laser of 532 nm was employed to record photoluminescence (PL) spectra of 1L-$WSe_2$ and Raman scattering spectra of 1L-$MoS_2$, as shown in Fig. 2f. A 100× objective lens was used to obtain the optical image and spectroscopic measurements.

Fig. S2a shows the PL spectra of $WSe_2$ fabricated by the robotic mechanical exfoliation system. A single prominent peak is observed at 1.67 eV in the PL spectra, which conclusively proves that the thickness of obtained $WSe_2$ is monolayer[1,2]. Fig. S2b presents the Raman scattering spectra of $MoS_2$ fabricated by the robotic mechanical exfoliation system. Two distinct Raman scattering peaks are observed at ~384 and ~403 $cm^{-1}$, corresponding to the in-plane $E^1_{2g}$ phonon mode and the out-of-plane $A_{1g}$ phonon mode, respectively[3]. The frequency difference between the two-modes is about 19 $cm^{-1}$, which suggests that the thickness of acquired $MoS_2$ is monolayer[4].

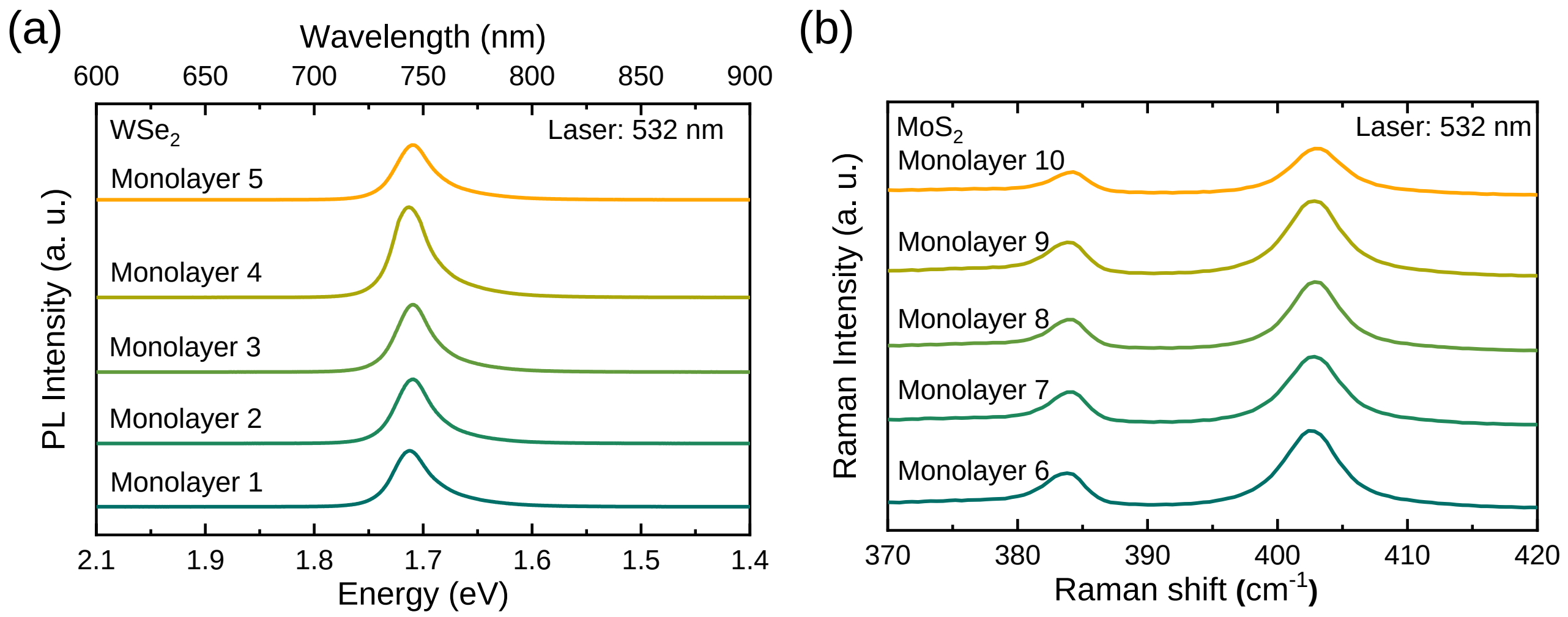

**Figure S3 Optical measurements of monolayer $WSe_2$ and $MoS_2$ fabricated by the automatic robotic system. (a)** Photoluminescence (PL) spectra of monolayer $WSe_2$ fabricated by the robotic system. **(b)** Raman scattering spectra of monolayer $MoS_2$ fabricated by the robotic system.

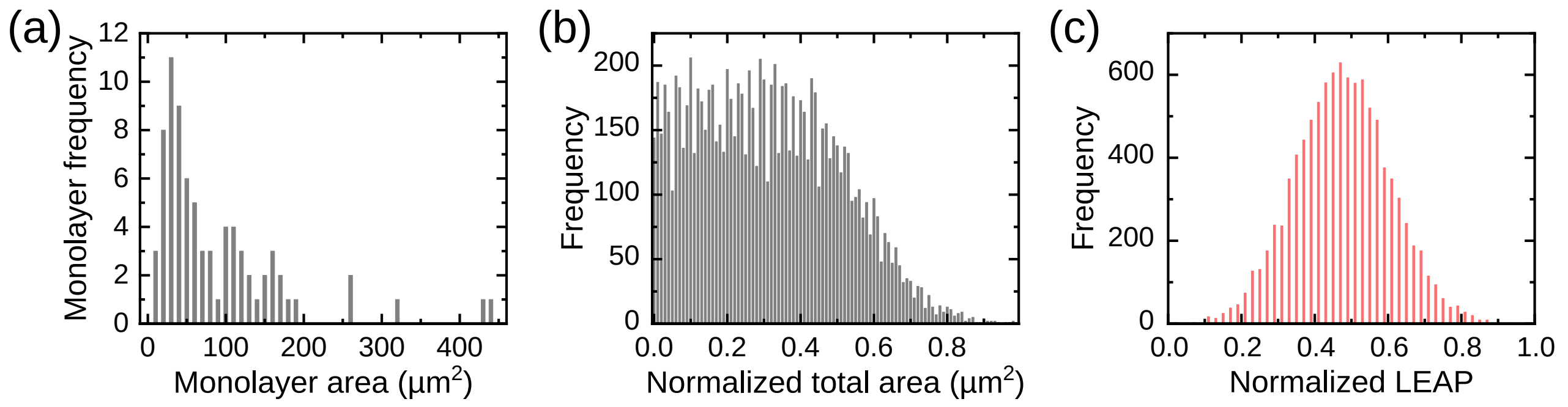

**Figure S4 Distribution of monolayer area in experiment, total simulation results of distribution of total area and LEAP.** (a) Averaged histogram of monolayer area obtained by the eight-times experiments conducted under the same conditions. (b) Frequency histogram of simulated total area. These are simulated by assuming that 100 monolayers are fabricated at a time according to a probability distribution from a normal-distribution of a hypothetical dataset with a maximum at 10 μm$^2$. (c) Simulated LEAP histrogram simulated by assuming that 100 monolayers are fabricated at a time according to a probability from a normal-distribution of a hypothetical dataset with a maximum at 10 μm$^2$.

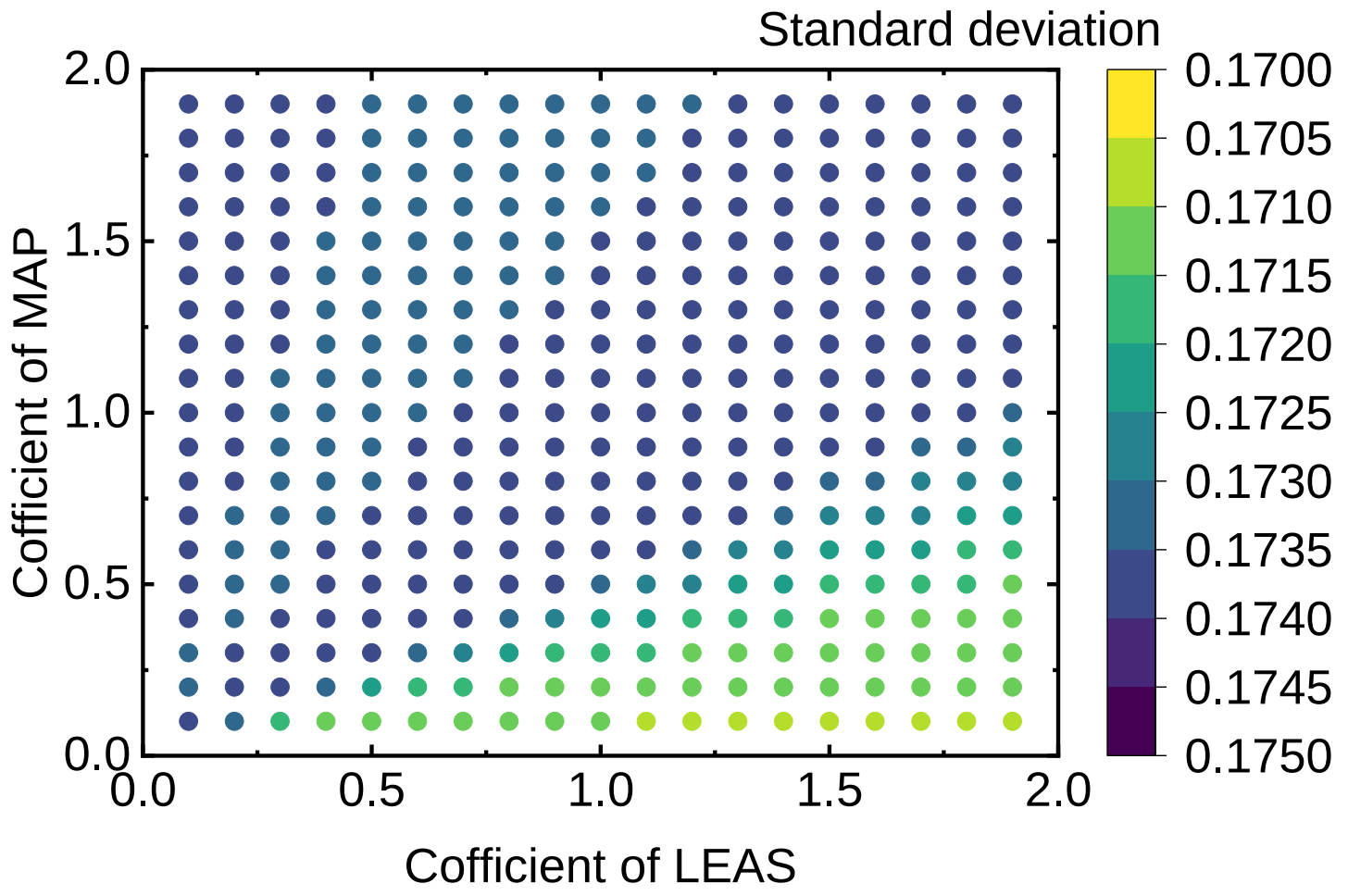

**Figure S5 Simulated standard deviation of LEAP as a function of MAP and LEAS coefficients $\boldsymbol{a_{1(2)}}$.** The results were simulated results assuming that 100 monolayers are fabricated at a time, following a probability with a normal distribution of the hypothetical dataset with a maximum at 10 μm$^2$. In the LEAP calcualtions, the standard deviation of LEAP is shown as a color map, when the MAP and LEAS coefficients are varied from 0 to 2. As the MAP coefficient decreases and the LEAS coefficient increases, the standard deviation of LEAP becomes smaller. However the difference of standrad deviation with varying the coeffcients of LEAP and LEAS is emough small within 0.5%.

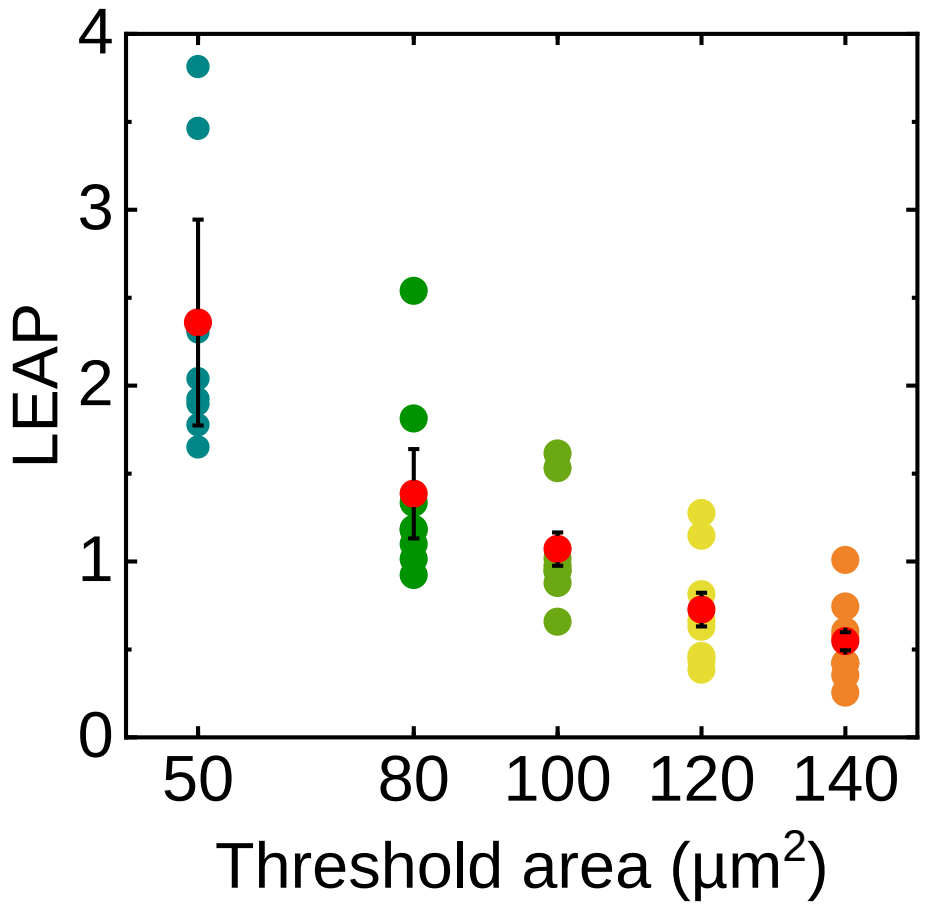

**Figure S6 Distribution of LEAP based on the experimental results of Fig. S4 with varying the monolayer threshold in the LEAS.** The results show that the standard deviation of LEAP becomes sufficiently small, when the threshold has the threshhold of exceedimng 100 μm$^2$.

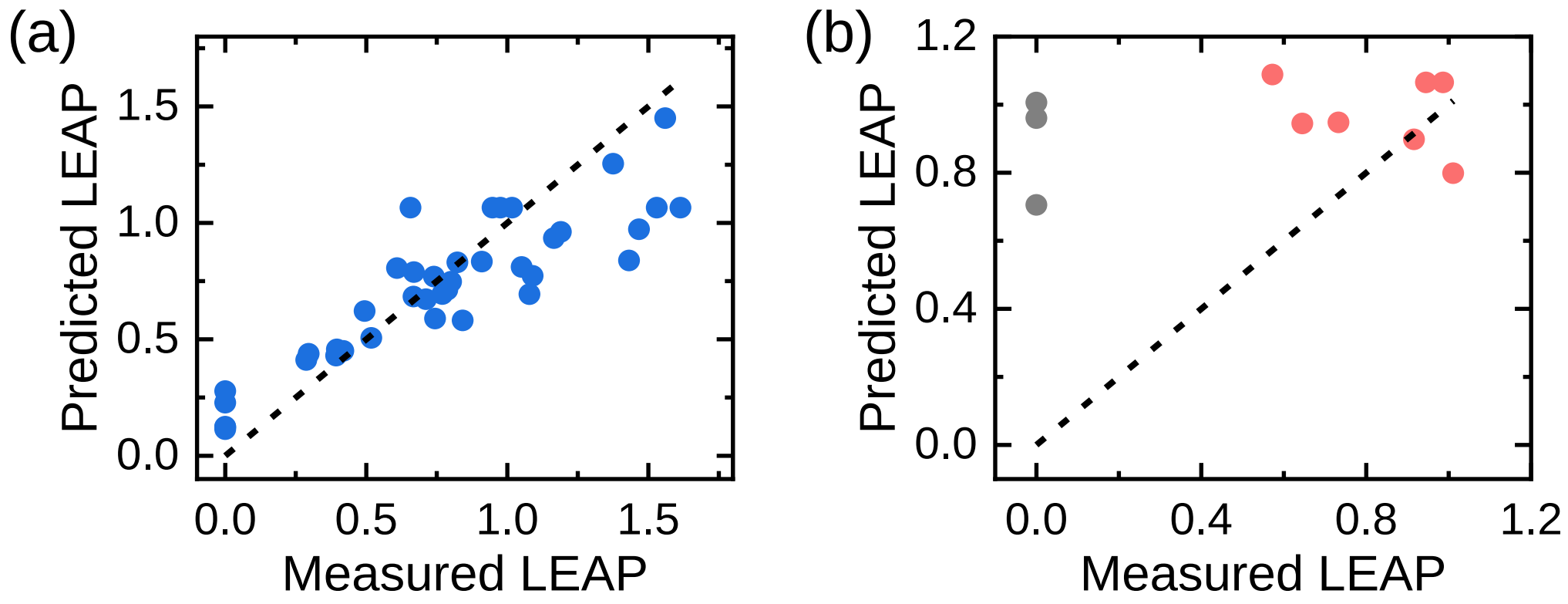

**Figure S7 Cross validation data for developing the model function.** (a) Cross-validation of experimentally measured LAEP and predicted LEAP by the model function constructed by Bayesian-optimization algorithm. The results of 20 experimental conditions conducted on the robotic system in advance were added to the training dataset. The datasets for 40 of 50 conditions were used as training data to construct the model function. (b) Cross-vlidation of 10 conditions, showing that the model function is more accurately predicted with the high LEAP values, as indicated by red circles. In the experiment, large deviation of simulated predicted LEAP values at measured LEAP around 0 as indicated by grey circles, suggests the experimental difficulties of the mechanical exfoliation method. This is because the bulk crystals may not be successfully exfoliated if the peeling velocity is extremely slow or the number of exfoliations is too small. Using this model function, we also performed the random simulations shown in Figure 4(a) and (b).

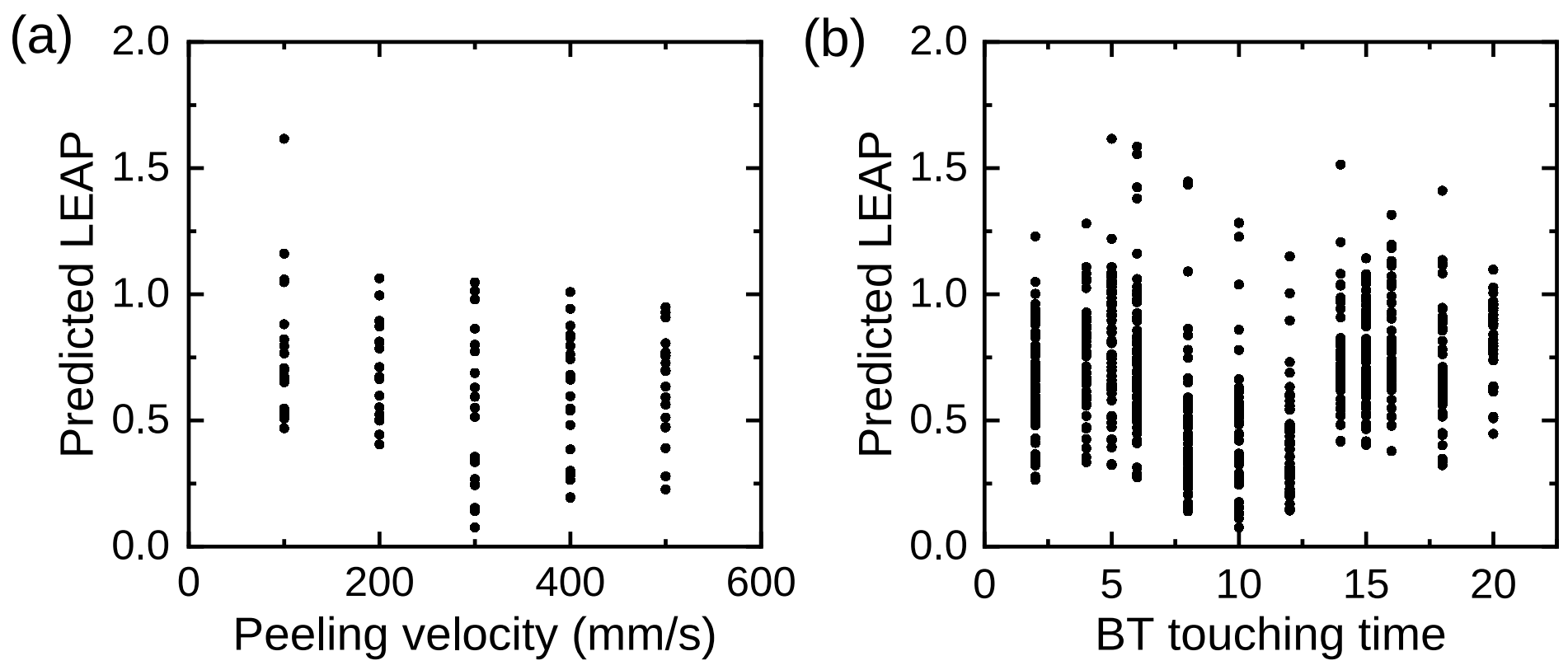

**Figure S8 Predicted LEAP as functions of (a) peeling velocity and (b) blue-tape touching time.** The predicted results were simulated for all experimental conditions using the model functions shown in Fig. S7. The two parameters of peeling velocity and blue-tape touching time, do not show a clear correlation with the maximum LEAP value.

| | Substrate material | Substrate thickness (μm) | Adhesive material | Adhesive thickness (μm) | Adhesion to silicon (N/20 mm) |
|---|---|---|---|---|---|
| BT 130 | EVA | 115 | Acrylic | 15 | 1.8 |
| BT 151 | EVA | 115 | Acrylic | 35 | 2.7 |
| BT 180 | EVA | 135 | Acrylic | 45 | 0.5 |
| BT 170 | EVA | 115 | Acrylic | 55 | 1.3 |
| BT 181 | EVA | 115 | Acrylic | 65 | 0.4 |

**Table S1 Detailed physical properties of blue-tape.** All blue-tapes used here are provided by Nitto Denko Corporation.